# The Case of Significant Variations in Gold-Green and Black Open Access: *Evidence from Indian Research Output*

Vivek Kumar Singh[a], Rajesh Piryani[b] & Satya Swarup Srichandan[a]
[a] Department of Computer Science, Banaras Hindu University, Varanasi-221005 (India)
Email: vivek@bhu.ac.in , satyaswarup98@gmail.com
[b] Department of Computer Science, South Asian University, New Delhi-110021 (India)
Email: rajesh.piryani@gmail.com

## Abstract

Open Access has emerged as an important movement worldwide during the last decade. There are several initiatives now that persuade researchers to publish in open access journals and to archive their pre- or post-print versions of papers in repositories. Institutions and funding agencies are also promoting ways to make research outputs available as open access. This paper looks at open access levels and patterns in research output from India by computationally analyzing research publication data obtained from Web of Science for India for the last five years (2014-2018). The corresponding data from other connected platforms- Unpaywall and Sci-Hub- are also obtained and analyzed. The results obtained show that about 24% of research output from India, during last five years, is available in legal forms of open access as compared to world average of about 30%. More articles are available in gold open access as compared to green and bronze. On the contrary, more than 90% of the research output from India is available for free download in Sci-Hub. We also found disciplinary differentiation in open access, but surprisingly these patterns are different for gold-green and black open access forms. Sci-Hub appears to be complementing the legal gold-green open access for less covered disciplines in them. The central institutional repositories in India are found to have low volume of research papers deposited.

**Keywords:** Gold Open Access, Green Open Access, Open Access, Paywall, Scholarly Articles, Sci-Hub.

## Introduction

Open Access refers to an academic movement where publications are made available and accessible free of cost with minimum restrictions to different kinds of audience. The **B**udapest **O**pen **A**ccess **I**nitiative (BOAI)[1] , defined "Open Access" as "making the publications freely availability on public internet, permitting any user to read, download, copy, distribute, print, search or link to the full texts of these articles, crawl them for indexing, pass them as data to software, or use them for any other lawful purpose, without financial legal or technical barriers". Open access brings significant benefits to authors, readers, funders, institutions etc., and more so for developing and under-developed countries, where enough resources are not readily available with academicians to pay the article access charges.

[1] www.budapestopenaccessinitiative.org, accessed on 25th June 2019.

Open access now has multiple forms. There are journals that are completely open access and as a result all articles published in that journal become openly accessible to everyone. The Directory of Open Access Journals (DOAJ)[2] is a community-curated online directory started in the year 2003 that indexes such journals. Then, there are journals which are usually behind a paywall for access, but make some articles accessible free to everyone, either after payment of an article processing charge or on their own after a particular period from the date of publication. Based on levels and types of open access, articles can be classified into following access categories (Piwowar et al., 2018):

Gold open access: This refers to an article which is published in a journal that is open access, i.e. it is in a journal where all articles are accessible freely to everyone. Usually open access journals are included in DOAJ.

Green open access: This refers to an article published in a journal that is behind a paywall but the article is uploaded on some repository (either disciplinary or institutional), which makes it accessible free of cost. However, in this case the reuse rights are restricted. Further, journals often impose an embargo period (varying from 6 to 48 months), after which the article can be uploaded on such a repository.

Hybrid open access: This refers to an article published in a closed journal which makes the article open access in exchange of an article processing charge paid by authors or their institutions through some agreement with the journal.

Bronze open access: This refers to an article that is free to read on the journal page but without a license to reuse.

Black open access: This refers to an article that is shared on illegal pirate sites, such as Sci-Hub or LibGen. However, this type is not well recognized as open access in the literature.

Closed access: This refers to all other articles that are not openly accessible in legal forms. This includes articles that may be shared on some academic social network or some illegal pirate sites.

Open access initiative practically began in 1991 when Paul Ginsparg of Los Alamos National Laboratory (LANL) setup arXiv[3], an electronic preprint service for Physics community. The invention of World Wide Web provided the right platform and infrastructure to make open access really possible. However, open access was first defined systematically in 2002 Budapest initiative, which itself was a result of meeting convened by Open Society Institute (OSI)[4] in 2001. This was followed by a series of other statements and declarations, including Bethesda Statement[5] and Berlin Declaration[6], both in 2003. Open access movement kept on gaining momentum thereafter, picking up pace during last 15 years, because of multiple reasons.

Many research funding agencies are now making it mandatory for grant recipients to publish their research outcomes in open access versions. Some of these agencies include US National Institutes of Health, the European Commission, US National Science Foundation, Welcome Trust, Bill and Melinda Gates Foundation etc. Several disciplinary repositories have emerged during this period that allow authors to freely upload their articles, sometimes the pre- or post-print versions. Authors are also uploading papers on academic social networking sites like ResearchGate and Academia. A good number of articles are also available for free download on pirate sites like Sci-Hub. Owing to different reasons, journals are also increasingly making many of their published articles openly accessible.

---

[2] Directory of open access (2018) Open access Repositories Operational Statues https://doaj.org
[3] https://arxiv.org/ accessed on 25th June 2019.
[4] https://en.wikipedia.org/wiki/Budapest_Open_Access_Initiative#cite_note-google.se-3 accessed on 25th June 2019.
[5] https://en.wikipedia.org/wiki/Bethesda_Statement_on_Open_Access_Publishing accessed on 25th June 2019.
[6] https://openaccess.mpg.de/Berlin-Declaration accessed on 25th June 2019

## Open access initiatives in India

The initial start of open access culture in India can be traced back to early 90s, when Indian physicists started depositing their preprints in arXiv. They were later joined by Mathematicians, Computer Scientists, Biologists etc. The Institute of Mathematical Sciences (IMSc) at Chennai actually setup a mirror server for arXiv. The Vidyanidhi Digital Library (Urs & Raghavan, 2001) from University of Mysore in 2002 is probably one of the first electronic thesis and dissertations initiative in India, a beginning of repository culture. The cause of public access to geographical information made in 1999 meeting of Indian Academy of Sciences[7], later published as a report in Current Science (Ramachandran, 2000), can be treated as one of the initial calls for public access to scientific information in India. The Indian Institute of Science, Bangalore setup EPrints@IISc electronic repository in 2002. By early 2002, several Indian institutions and organizations have started taking initial steps towards open access of various scientific information. The National Knowledge Commission in a report[8] in 2007 discussed about open access and open educational resources. In 2009, CSIR[9] headquarters sent a memorandum to all its 37 laboratories in the country to set up institutional open access repositories, though it took some time for them to start implementing that. The studies by Das (2008) and Arunachalam & Muthu (2011) present a good report on early open access initiatives and projects in India.

The more recent effort by the Government of India is the Open Access Policy[10], released in 2014, jointly by Department of Biotechnology (DBT) and Department of Science and Technology (DST). DST and DBT are the two main research departments of Ministry of Science and Technology of Government of India that fund different kinds of research activities. The policy states that "*since all funds disbursed by the DBT and DST are public funds, it is important that the information and knowledge generated through the use of these funds are made publicly available as soon as possible....*". The main guiding principle of this policy seems to be the fact that public funded research outcome should be publicly accessible. The policy also envisaged that open access will allow percolation of cutting-edge research in Higher Education curricula, which in turn will raise the standards of technical and scientific education in the country. Through the policy, institutions were encouraged to setup institutional repositories (IRs) and to deposit all research outcome coming out from them in the repositories. A central harvester *sciencecentral.in* was also created and it was expected that all institutional repositories will eventually link to it. The policy was mandatorily to be followed by institutions receiving core funding from DBT and DST.

The Council of Scientific and Industrial Research (CSIR), a major institutional system of scientific and industrial research in country, has issued an open access mandate[11] which requires all CSIR laboratories to setup their own interoperable institutional open access repositories. Through this mandate CSIR took initiative to "lead the open access movement within the country". The *csircentral.net* established by CSIR-URDIP is the central harvester that is expected to link different institutional repositories created by individual CSIR labs. The Indian Council of Agricultural Research (ICAR), the body guiding and promoting agricultural research in country, also issued an open access policy[12], which requires each ICAR institute to setup an open access institutional repository. ICAR has also setup a central harvester for one stop access to all the agricultural knowledge generated in ICAR.

[7] Indian Academy of Sciences https://www.ias.ac.in/ accessed on 20th August 2019
[8] National Knowledge Commission. (2009). National Knowledge Commission. Report to the Nation 2006-2009. http://14.139.60.153/bitstream/123456789/112/1/National%20Knowledge%20Commission-Report%202006-2009%20.pdf accessed on 20th Aug. 2019.
[9] http://www.csir.res.in accessed on 20th Aug. 2019
[10] DBT and DST Open Access Policy Policy on open access to DBT and DST funded research http://www.dst.gov.in/sites/default/files/APPROVED%20OPEN%20ACCESS%20POLICY-DBT%26DST%2812.12.2014%29_1.pdf accessed on 20-08-2019.
[11] http://www.csircentral.net/mandate.pdf

In 2018, a group of academicians and open access enthusiasts signed the Delhi Declaration on Open Access[13], which advocated for "the practice of open science" and "adoption of open technologies for the development of models for sharing science and scholarship". The more recent National Digital Library project of Ministry of Human Resources and Development of Government of India tried to bring the repository culture to Higher Education Institutions of India. The IndiaRxiv[14] of the Open Science project is the most recent addition to the open access cause. Open Access India has initially launched AgriXiv- a preprints repository for agriculture and allied sciences- and then launched IndiaRxiv. However, this preprints repository is yet to pick up momentum as there are only 67 preprints available as on 16th Feb. 2020, from the time of launch of the repository in April 2019.

## Research Questions

This paper aims to explore and analyze the open access levels and patterns in research output from India during last five years, i.e. 2014 to 2018. More precisely, it aims to answer following research questions:

**RQ1**: Do there exist significant variations in volume of Indian research papers available in different forms of open access?

**RQ2:** How much research output could be accessed through the major Institutional Repositories in India?

**RQ3:** How does open access levels and patterns vary across disciplines?

**RQ4:** What may be plausible reasons for observed low volume in gold and green open access as compared to high volume observed in black open access?

## Related Work

There are some recent studies that tried to measure the open access levels in research outputs and also to characterize them with respect to time, publisher or discipline. Some studies focused on specific countries/ regions and few others on particular subjects. Different studies used different kinds of data sources. Here we present some of the most relevant studies to the present work. We did not find any study that looks at all types of open access, with most either exploring gold and green or only black open access. at the paragraphs below mention some previous studies that explored either the legal gold & green open access or the black open access through Sci-Hub.

---

[12] https://icar.org.in/hi/node/5542
[13] http://openaccessindia.org/delhi-declaration-on-open-access/
[14] https://indiarxiv.org/

A relatively old study (Hajjem et al., 2005) analyzed the impact measures of growing OA and developed a robot that can crawl the web for 1,307,038 full-text articles for 12 years (1992-2003) in 10 disciplines using reference metadata (author, title, journal, etc.) and citation data from the Institute for Scientific Information (ISI) database. They also found that OA articles have comparatively more citations than non-OA, varying from 36%-172%.

Bjork et al. (2010, 2012, 2014, 2016, 2017) through their multiple studies during 2010 to 2017, estimated and characterized the open access levels and patterns in scientific publishing. They mainly focused on analyzing open access levels (Gold, Green, Hybrid) of scholarly articles through various models and studies such as hybrid model for open access, longitudinal studies of open access, and anatomy of green open access etc.

Archambault et al. (2013, 2014) in their reports analyzed the proportion of open access peer-reviewed papers at European and world levels for 2004-2011 and 1996-2013 time periods, respectively. They have shown that tipping point for open access (more than 50% of the research articles are freely available) has been attained in several countries, including Brazil, Switzerland, Netherlands, US, as well as in biomedical research, biology, and mathematics and statistics.

Piwowar et al. (2018) used three different samples of 100K articles each drawn from Web of Science (WoS), CrossRef and Unpaywall for publications from worldwide. Based on analysis of the samples, they estimated that about 28% the scholarly articles are open access. This accounts for about 19 million of 67 million total articles used in the estimate. They found significant annual growth in open access levels, particularly in Gold and Hybrid, reaching as high as 45% for the year 2015. They also analyzed citation levels of open access articles, concluding that open access articles attain about 18% more citations than average.

Bosman & Kramer (2018) in a recent study collected data from Web of Science using its oaDOI service and explored open access levels across research fields, languages, countries, institutions, funders and topics. They found high variations in open access levels on all these dimensions. With respect to countries, they found open access levels varying from 10% to 60%. Further, with respect to disciplines, they found higher open access levels (more than 50%) in disciplines like Life Sciences, Biomedicine and Physical Sciences as compared to relatively lower levels (less than 20%) in Social Sciences and Arts & Humanities.

Martin-Martin et al. (2018) analyzed a large sample of publication data from Web of Science, obtaining the open access evidence from Google Scholar. Articles indexed in Web of Science for year 2009 or 2014 (all articles with a DOI in SCI, SSCI, and A&HCI) were analyzed by country and discipline, differentiating by types of OA. Among the articles that were analyzed there were over 50,000 articles published in 2014 from Indian researchers, and the analysis showed that about 23% of research output from India was available in legal forms of open access. They also showed that that ResearchGate provided access to 34% of the articles from India.

Several previous studies aimed to explore the volume of scholarly literature available in Sci-Hub, which is often regarded as black open access. One of the most comprehensive studies on Sci-Hub is by Himmelstein et al. (2017). This study observed that Sci-Hub has grown rapidly since its creation in 2011 and as in March 2017, it contained 68.9% of the 81.6 million scholarly articles registered with Crossref and 85.1% of articles published in toll access journals. The study found some variations in Sci-Hub coverage of articles with respect to disciplines and publisher and concluded that Sci-Hub provides access to nearly all scholarly literature. Cabanac (2015) is one of the initial studies that discussed about *biblioleaks* as in the case of Sci-Hub. Bohannon (2016a, 2016b) reported about Alexandara Elbakyan, the founder of Sci-Hub and her version of the objectives of Sci-Hub, and worked with her to obtain access log of Sci-Hub (Elbakyan & Bohannon, 2016). He analyzed the access log and reported his findings about who is downloading pirated papers on Sci-Hub (Bohannon, 2016b). He concluded that Sci-Hub download activity is spread across the world, irrespective of availability or non-availability of legal access, with a good amount of download coming from institutions that already have legal access to downloaded articles. Elbakyan (2016a, 2016b) explains that Sci-Hub is a goal and claimed that it is a "true solution to open access". Few other studies (Faust, 2016; Greshake, 2017; Priego, 2016) explored the Sci-Hub platform and the probable reasons for its widespread use and argued that Sci-Hub is at least a signal of failure of current open access models if not a solution to open access. It was also observed that Sci-Hub is trumping gold and green open access and that a change in approach for open access is needed (Green, 2017). Strielkowski (2017) explored the question whether rise of Sci-Hub will pave the road for subscription-based access to publishing and suggested a subscription-based model as a solution to open access. Travis (2016) observed that most people give "thumbs-up" to pirated papers. Novo & Onishi (2017) even went to ahead to argue if Sci-Hub could become a quicksand for authors. Some studies looked at Sci-Hub with reference to specific geography of Latin America (Machin-Mastromatteo et al., 2016; Mejia et al., 2017). No previous studies, however has been found on exploring Sci-Hub about research output from India and to what extent Sci-Hub provides access to Indian research output.

Among studies mainly focusing on open access in India, (Bhardwaj, 2015) looked at India's contribution to open access movement, whereas (Madhan et al., 2017) presented a discussion on whether Indian authors should pay to publish their work as open access. The recent study by Piryani et al. (2019) tried to analyze the overall level of open access in India by taking data from web of Science for all publications during 2016 that have at least one Indian author. They conclude that overall open access level in Indian research output is about 24% of the total output, which is less than the world average. However, this study only looked at India as a county overall and did not analyze the data in black open access or in Institutional Repositories in India. Some other studies focused on Institutional Repositories in India that provide Green open access to research articles. Kumar & Mahesh (2017) tried to analyze the Institutional Repositories in India, as a means of providing open access to articles. They observed the statistics of submission of papers for about one year and found that submissions to the Institutional Repositories was very low, some not even getting even a single paper deposited in the whole year. They concluded that the Institutional Repositories in India have not really picked up in terms of papers deposited. Several other studies (Ghosh et al., 2007; Nazim & Devi, 2008, Singh, 2016; Roy et al., 2016; Momin & Gaonkar, 2016; and Panda, 2016) focused on different aspects of Institutional Repositories ranging from publication volumes & impact to deterrents to submissions in Institutional Repositories in India. A recent article in Nature (Mallapaty, 2019) talked about the newly proposed IndiaRxiv repository, with the note that performance of Institutional Repositories in India is not very good.

To the best of our understanding, there are no previous studies that analyze all the three forms- gold, green and black open access- of research articles, more so for a country like India. This paper tries to bridge this gap and presents a comprehensive study on open access levels in all forms for Indian research output, along with analyzing disciplinary variations in open access levels.

## Data and Methodology

The core data for the study is the research publication data from India during the period 2014 to 2018, obtained from Web of Science (WoS) index. The data was obtained from the three main citation indices of Web of Science, namely Science Citation Index Expanded (SCIE), Social Science Citation Index (SSCI) and Arts and Humanities Citation Index (AHCI). The data was obtained through query on country filed tag (CU) of Web of Science. The publication year range was set by using the PY tag of Web of Science, which was set to 2014 - 2018. Thus, the query helped in obtaining all research output indexed in Web of Science for the years 2014 to 2018, with at least one author affiliation from India. A total of 377,336 publication records were found. Out of these 377,336 records, only 335,503 records had a DOI. Since we wanted to obtain information about the papers being analyzed, from different platforms (such as Unpaywall and Sci-Hub), we needed a linking information for the same paper across different platforms. The DOI was the most appropriate thing for this purpose. Therefore, publication records not having DOI were dropped from further analysis and only the 335,503 publication records having DOI were analyzed further. The data was downloaded in the month June 2019.

In order to find out what proportion of data downloaded is open access (gold or green), the Unpaywall REST API service was used. For each publication record in the data, the DOI was used to lookup information about the paper from Unpaywall. The data from Unpaywall website was retrieved during 27-29 June 2019. The Unpaywall REST API returns data in JSON format, which contain information like open access status, host type etc. The Unpaywall data for WoS records was obtained through a program written in Python. It was found that Unpaywall had information for 332,111 out of the 335,503 records.

The third step in data collection was to obtain relevant data from Sci-Hub for each of the Web of Science records being analyzed. This was done by writing an automated querying program which looked for availability of full-text of research paper. However, since obtaining information about all articles being analyzed would have taken a long duration, we limited lookup in Sci-Hub only for publication records for the year 2016. Thus, the automated lookup in Sci-Hub platform was made for 67,857 publication records with DOI for the publication year 2016.

The fourth part of data collection was to obtain the statistics for volume of research output accessible through the major Indian Institutional Repositories (IRs). It may, however, be noted that there was no way to programmatically find out which papers are available in Indian IRs. Therefore, we simply obtained counts of research papers available in the IRs. Counts for the three main central IRs, namely *sciencecentral*, *csircentral* and *krishikosh* were obtained. The numbers were analyzed in the context of publication records from India indexed in Web of Science.

Since the paper aimed at characterizing open access levels in different disciplines, each publication record was tagged into one of the 14 broad research disciplines. This disciplinary categorization into 14 broad disciplines was proposed in a previous work (Rupika et al., 2016) and used thereafter by many publications. The WoS Category (WC) field was used for tagging each publication into disciplinary categories. These 14 broad research disciplines in which articles were tagged are: Agriculture (AGR), Art & Humanities (AH), Biology (BIO), Chemistry (CHE), Engineering (ENG), Environment Science (ENV), Geology (GEO), Information Sciences (INF), Material Science (MAR), Mathematics (MAT), Medical Science (MED), Multidisciplinary (MUL), Physics (PHY) and Social Science (SS). The disciplinary characterization of the research output was done by using these tags assigned to each publication record. Different programs were written in Python to computationally process the whole data and produce analytical results. The results obtained are shown in different tables and figures.

## Results

The sub-sections below present analytical results about volume of Indian research output available in recognized legal open access forms, black open access through Sci-Hub and also statistics for Indian central IRs.

### *Gold and Green open access as observed in publication records from Web of Science*

The first point of analysis was to find out what proportion of research output from India for last five years (2014 – 2018) is openly accessible as gold or green open access. Results show that out of 335,503 scholarly articles with DOI in WoS, a total of 332,111 records were captured by Unpaywall. Out of these records, a total of 80,331 (**24.19%**) records were available in open access. This finding agrees with earlier findings of Martin-Martin et al. (2018), which found that about 23% of articles from India for the year 2014 are available in legal open access forms. **Table 1** shows the detailed statistics for all the years from 2014 to 2018. It can be observed that there is continuous growth in number of articles that are open access, though proportion wise its approximately same level throughout all the years (with a little growth seen in 2016). It would be interesting to compare this with the world average. As reported in previous studies for the years 2015 and 2016, approximately 30% of articles for the world are open access. In case of India this level is around 24% of the total research output. Thus, India's open access proportion in its total output is a bit lower than the world average.

The second analysis was performed by categorizing articles accessible in different types of open access, i.e. gold, green, bronze, and hybrid, except black open access. **Figure 1** shows the proportion of articles from India that are openly accessible and the type of open access. It is observed that gold open access is the most common type of open access among legal open access models, with about 10%-12% articles being gold open access. Green open access is the second common type of open access, with about 6% of articles being green open access. This is followed by bronze (around 5%) and hybrid (around 3%). Here, it may be noted that the numbers for all these open access forms are non-overlapping, i.e. green open access refers to articles that are available as green open access only and not in any other form. Similarly, gold open access is only for exclusive gold open access articles and does not include articles with any other open access type. Gold open access has increased in almost all years, except 2017 and 2018. The slight drop in open access levels in recent two years can be explained from the fact that many of the recently published articles may still not be available as open access due to several reasons including embargo periods imposed by journals.

For a more detailed understanding of open access patterns in India, we further analyzed the research output data for the 100 most productive institutions. The number of research papers produced and the amount of openly accessible papers for each institution was computed. **Figure 2** shows these institutions on a plot of total number of papers vs. number of papers that are openly accessible. It is observed that institutions under CSIR system, IISc Bangalore, AIIMS, TIFR etc. have higher proportion of their output available in open access. It was also observed that main discipline of research of an institution had a close connection with open access levels. Institutions specializing in Medical Science, Physics etc. observed higher open access levels as compared to others. No significant differentiation in open access level was, however, seen with respect to geographical location (big city, urban area or rural area) of an institution.

### *Black open access from Sci-Hub*

As we observed in the previous section, the volume of papers available in gold and green open access for Indian research is found quite low, with less than 1/3rd of total papers being openly accessible. In this section we look at what proportion of the Indian research output was available for free download in Sci-Hub. The Sci-Hub website on being queried through the automated program for the 67,857 publication records with DOI for the year 2016, resulted in a total of 61,706 articles being accessible with full text. This constitutes about 91% of the total research output as against only 24% research output being openly accessible in legal open access forms (such as gold, green open access). **Table 2** presents the relevant data about articles accessible in Sci-Hub.

We were also interested in knowing as to what amount of these Indian articles for the year 2016 available in Sci-Hub is being downloaded. For this purpose, we downloaded the Sci-Hub access log for 2017 (Tzovaras, 2018). We parsed the access log to extract all entries for Indian papers for the 2016. We found that out of total of 28,773,939 total articles downloaded, a total of 36,333 papers downloaded are the papers from India published in 2016 and indexed in Web of Science, i.e. our data sample. These papers taken together are downloaded a total of 480,924 times, i.e. average download per paper is 13.23. **Figure 3** shows the download activity of Indian papers for 2016 in the calendar year 2017. It can be observed that on an average 2,500 downloads happen every day, with 1,500 of them being unique downloads. However, in absence of any previous studies on this for India or other geographies, it is not possible to compare the value observed with other results and to understand if it is a low or a high value.

The next step was to try to find out where are the downloads to Indian papers coming from. Are these download requests to Indian papers in Sci-Hub coming mainly from India or they are more wide-spread geographically? In order to analyze this, we processed the complete access log to extract the IP address and country information of the persons downloading the articles. This information is then geotagged and plotted on world map. **Figure 4** shows the origin of download activity during 2017 of the Indian papers published in 2016 as available in Sci-Hub. It is interesting to observe that the downloads are originating from all across the world. In fact, most intensive download activity is from Northern Europe, where institutions may have one of the richest journal subscriptions. A significant number of downloads are also originating from United States, Mexico, South America and South east Asian countries, China and Japan. Thus, it appears that Sci-Hub is a preferred source for article download, possibly because of its simplicity, even if there are legal access routes to articles.

### *Open access through Institutional Repositories*

Several organizations in India have tried to promote open access in variety of ways, including by setting up Institutional Repositories. The DST-DBT's *sciencecentral.in*[15], CSIR's *csircentral.net*[16] and ICAR's *krishikosh*[17] are some of the most prominent central institutional repositories in the country. Several government organizations are now making it mandatory for their scientists and researchers to deposit their research output in relevant IRs. Research funding agencies are also increasingly calling for public access to public funded research and encouraging the grantees to submit their research outcomes in relevant IRs. Thus, IRs can play a major role in open access of research articles. We tried to analyze the number of articles available through some of the important IRs, as detailed below.

[15] http://www.sciencecentral.in/
[16] http://csircentral.net/
[17] https://krishikosh.egranth.ac.in/

The three main central repositories in India are DST-DBT's *sciencecentral.in*, CSIR's *csircentral.net* and ICAR system's *krishikosh* repository. All these central repositories taken together have 158 Institutional Repositories connected with them. **Table 3** shows the detailed data about number of articles available in each of these three main central repositories.

DST-DBT's *sciencecentral.in* provides harvester service that harvests the full text and metadata of the publications from DST-DBT institutions and funded research. The website for the IR says "*at present there are 17 Institutional Repositories hosted at the science central, while 42 institutional repositories are regularly harvested on the same*". We found that there are 125,595 total publications deposited and available through the 24 Institutional Repositories that are part of the central repository. Even if assume that all submitted papers are indexed in Web of Science, this constitute only 15% of total papers indexed in Web of Science for the concerned period (8,09,404). This number is the total number of publications from India indexed in Web of Science. However, it is quite clear that all the deposited papers will not all be indexed in Web of Science, therefore the percentage of research papers actually accessible through this IR will be much lesser than 15%.

The *csircentral.net* repository connects about 30 IR's with approximately 100,609 articles deposited. In this case too, even assuming all papers deposited in the IR are indexed in Web of Science, the papers accessible through IR will be less than 13% (with 747,759 papers from India indexed in Web of Science during 2006-2019 period). In reality the papers deposited will not all be indexed in Web of Science and hence the actual percentage of papers accessible through this IR would be much lesser than 13%.

The *krishikosh* repository, known in full as Knowledge based Resources Information Systems Hub for Innovations (KRISHI) in agriculture, is a centralized data repository system of ICAR consisting of Experiments/ Surveys/ Observational studies, Geo-spatial data, Publications, Learning Resources etc. The KRISHI community is divided into Subject Matter Division (SMD) such as Agricultural Education (1285), Agricultural Engineering (364), Agricultural Extension (235), Animal Science (1720), Crop Science (4025), Fisheries (2721), Horticulture Science (3677), Natural Resource Management (4252) etc. The numbers indicated in brackets are the number of resources available in repository. A total of 18,486 records are available as on date in all connected repositories taken together for *krishikosh*. In this case too, the number of papers that are accessible through IRs is very low as compared to publication volumes.

Another recent initiative in the repository way of open access in India is the creation of IndiaRxiv hosted at indiarxiv.org. However, this has yet to pick up as at the moment it has only 67 publications uploaded as on 16$^{th}$ Feb. 2020, from the time of launch in April 2019. Perhaps it will take significantly more time for this platform to get good number of articles deposited.

As can be observed from discussion above, it is quite clear that the amount of papers deposited in Indian IRs is much lesser than publication volume of the Indian institutions. The analysis of the three major repositories: *sciencentral*, *csircentral* and *krishikosh* suggest that more efforts are needed to promote the IR culture in India. These efforts could not only be limited to funding agencies but all the research institutions should make it mandatory for their researchers to submit their papers (or pre-print or post-print) to concerned IRs. Individual scientists should also be encouraged and incentivized for submitting their papers to IRs. This would not only help other Indian researchers who do not have access to costly journal subscriptions but also the researchers themselves as their paper would find more citations and use.

### *Disciplinary Variations in Open Access*

In addition to analyzing the overall level research output available in open access, the paper also tried to analyze disciplinary variations in open access levels of Indian research output. **Table 4** shows the number of research articles from India for each discipline produced in 2016. It also shows number of articles for each discipline that are open access, either legal forms or in black open access. It is observed that only Multidisciplinary, Medical Science and Physics disciplines have more than 40% of their articles available in legal open access, either gold or green. Physics and Mathematics are the other disciplines with close to 40% of the articles available as open access. Arts & Humanities, Information Science and Chemistry disciplines have lesser proportion of articles available in open access.

The analysis of availability of articles for 2016 in Sci-Hub show that black open access in all disciplines is much higher, ranging from 78% for multidisciplinary field to 97% for Chemistry. Nine out of fourteen disciplines have more than 90% of the articles available in Sci-Hub for free download. These disciplines are Chemistry, Engineering, Environmental Science, Material Science, Physics, Biological Science, Information Science, Social Science and Arts & Humanities. It is interesting to observe that disciplines which have less percentage of articles available as gold and green open access are one of the highest covered disciplines in Sci-Hub. It can be clearly seen that disciplines like Arts & Humanities, Information Science, Engineering and Engineering that have the lowest percentage of articles available in gold and green open access, have more than 90% of articles available for free download in Sci-Hub. The lack of well-developed disciplinary repositories in Arts & Humanities and Social Sciences may be one of the primary reasons for low volume of research papers from these disciplines being available in legal open access forms. However, Sci-Hub seems to be filling this gap with higher coverage in these disciplines. In this sense, Sci-Hub is seen complementing the low open access availability in gold and green forms in some disciplines.

## Summary & Conclusion

The paper tried to measure and characterize open access levels and patterns in research output from India, with special focus on differences in volume of articles available in different forms of open access. Looking at differences in legal forms of open access (such as gold & green) and black open access through Sci-Hub was one of the interesting question being explored. Disciplinary variations in open access levels in different open access forms was also analyzed. Interesting results are obtained from the analytical study. The key findings, as far as they answer the research questions proposed, are as follows:

***Firstly***, it is found that only about 24% articles from India published during last five years are available as legal open access. This is about 6% lesser than the world average. Gold open access is found to be more prevalent than green open access. However, gold and green open access taken together provide open access to less than 1/3$^{rd}$ of research papers from India. On the contrary, Sci-Hub provides free access to more than 90% of research papers from India, as observed for 2016 data. Thus, there exist a huge difference in volume of research papers available as open access in legal gold & green forms and the black open access provided by Sci-Hub. It may be noted that Sci-Hb is not a well-recognized legal open access system and that it has faced many problems, including several legal suites. However, availability of significant amount of research output in Sci-Hub and good number of downloads from the platforms from across the world, is a clear implication that traditional and legal open access models have failed to truly deliver open access to research articles.

***Secondly***, the analysis of data from the three main central Institutional repositories in India show that a very low volume of research output from India is getting deposited in Institutional Repositories. Estimates show that this is less than 15% of the research output produced from India. This implies that India has somehow failed to tap the potential of Institutional repositories for providing open access to its research output. It appears that open access initiatives should not be limited to mandates by research funding agencies, rather much more needs to be done for promoting open access through Institutional Repositories. Perhaps all the research institutions should make it mandatory for their researchers to submit their papers (or pre-print or post-print) to concerned IRs. A more productive approach could be to provide incentives to individual scientists as well as institutions that submit most of their papers to Institutional Repositories.

***Thirdly***, the discipline-wise analysis shows that there exist disciplinary variations in levels of open access. However, interestingly these patterns are found to be different for legal gold & green open access and black open access through Sci-Hub. While, disciplines like Multidisciplinary, Medical Science and Physics have higher proportion of articles available in gold & green open access; the Multidisciplinary and Medical Science disciplines are not the best covered in Sci-Hub. Disciplines like Chemistry, Information Science and Arts & Humanities that have less percentage of research papers available as gold & green open access have very high coverage in Sci-Hub with more than 90% papers available for free download. This is a very interesting pattern observed. It appears that Sci-Hub is in a sense complementing the gold & green open access models, with preferential coverage of disciplines that are not well-covered through legal gold & green open access models.

***Fourthly***, it is clearly observed that despite several initiatives and a push from funding agencies and other bodies, the amount of papers accessible through legal gold & green open access models is significantly much lesser as compared to black open access. There could be several reasons for this pattern, as far as it relates to India. one of the main reasons could be the fact that Indian researchers are not really taking the open access calls seriously. Though lower volume of gold open access is not in control of researchers but volume of papers accessible through green open access can certainly be many times higher than the current levels. More efforts and perhaps incentive mechanisms may be needed to persuade researchers to deposit their papers (including pre- or post-prints) to Institutional or disciplinary repositories (like arXiv) to make the papers openly accessible. Further, Indian institutional system seems to have lack of funds to enter into open access agreements with different journals and publishing houses to make research papers openly accessible. Given that many developed countries are actively considering joining Plan-S, it would be desirable that Indian science administrators take a view on this. Another important point to note is the high volume of papers available in Sci-Hub as well as number and geographies of downloads. As observed in previous sections that a good number of downloads from Sci-Hub are originating from places that usually have rich access to research papers through legal forms. Perhaps the simple, straightforward and easy to use system of Sci-Hub is tempting researchers to use it's a preferred or reliable source to look for research papers. Further, researchers downloading papers from Sci-Hub seem totally unconcerned with legal issues associated with Sci-Hub's model. This observation has at least once very clear implication that the access models and platforms have to be simplified and made easy to use.

The current research work could not look into some other important aspects of open access of Indian research output, that could be taken up as future work. *First*, it does not analyze the amount of research papers from India deposited in Disciplinary repositories. Disciplinary repositories being preferred sources for authors, it would be interesting to analyze as to what amount of research output for different disciplines from India is available in open access through relevant disciplinary repositories. *Second*, the paper could not analyze the data from academic social networks like ResearchGate that now have a good amount of research papers available from different countries. Irrespective of copyright issues, these platforms provide open access to a good volume of research papers. Mainly absence of any APIs and restrictions on automated crawling volumes permitted on the ResearchGate website prevented us from including this analysis in current work. It would, however, be a very interesting study to look at the amount of research output from India available in academic social networks such as ResearchGate.

## Acknowledgments

The authors acknowledge the enabling support provided by the DST-NSTMIS funded project '*Design of a Computational Framework for Discipline-wise and Thematic Mapping of Research Performance of Indian Higher Education Institutions (HEIs)*', bearing Grant No. DST/NSTMIS/05/04/2019-20, for this work.

# Tables

**Table 1: Research Output from India indexed in WoS (2014-2018) that are legally OA**

| Publication Year | No. of Articles indexed in WoS | No. of Articles in WoS having a DOI | No. of Articles found on Unpaywall# | No. of articles that are OA# |
|---|---|---|---|---|
| 2014 | 67,575 | 58,310 | 58,308 (99.99%) | 13,834 (23.72%) |
| 2015 | 70,685 | 61,695 | 60,039 (97.32%) | 14,347 (23.89%) |
| 2016 | 76,530 | 67,857 | 67,370 (99.28%) | 17,280 (25.64%) |
| 2017 | 78,532 | 70,861 | 69,989 (98.77%) | 17,014 (24.30%) |
| 2018 | 84,014 | 76,780 | 76,405 (99.51%) | 17,856 (23.37%) |
| **Total** | **377,336** | **335,503** | **332,111 (98.99%)** | **80,331 (24.19 %)** |

#- Percentage is calculated with respect to number of articles in WoS having a DOI and found in Unpaywall, i.e. column 4.

**Table 2: Articles in WoS from India that are on Sci-Hub**

| No. of articles in WoS (PY=2016) | Articles with DOI | No. of articles that are on Sci-Hub | Sci-Hub articles as Percentage of total articles[1] |
|---|---|---|---|
| 76,530 | 67,857 | 61,706 | 90.93 |

[1] Percentage is calculated with respect to number of articles in WoS having a DOI, i.e. column 2.

**Table 3: Status of the main/ central Institutional Repositories in India**

| S. No. | Name of Central Repository | No. of Connected IRs | Period of Records | Total Records |
|---|---|---|---|---|
| 1 | DST-DBT Central Repository (sciencecentral.in) | 24 | 1920-2019 | 125,595 |
| 2 | CSIR Institutional Repository (csircentral.net) | 30 | 2006-2019 | 100,609 |
| 3 | ICAR Central Repository (krishikosh) | 104 | NA | 18,486 |

**Table 4: Discipline-wise Distribution of Articles that are Gold-Green OA or available in Sci-Hub**

| Discipline | No. of Article in WoS (PY=2016) | No. of Articles with DOI | Articles that are available as gold and green OA | | Articles that are on Sci-Hub | |
|---|---|---|---|---|---|---|
| | | | No. of Articles that are OA | Percentage | No. of Articles that are on SCI-HUB | Percentage of articles available on SCI_HUB[1] |
| MED | 15,307 | 11,884 | 6,118 | 51.48% | 9,422 | 79.28% |
| CHE | 14,092 | 13,496 | 1,337 | 9.90% | 13092 | 97.01% |
| PHY | 7,838 | 7,598 | 3,411 | 44.89% | 7,286 | 95.89% |
| BIO | 7,094 | 6,394 | 2,525 | 39.49% | 5,871 | 91.82% |
| ENG | 6,708 | 6,389 | 957 | 14.97% | 6,120 | 95.79% |
| MAR | 5,818 | 5,562 | 1,110 | 19.95% | 5,356 | 96.3% |
| AGR | 4,582 | 3,516 | 1,245 | 35.40% | 2,803 | 79.72% |
| GEO | 3,912 | 3,038 | 916 | 30.15% | 2,628 | 86.5% |
| ENV | 3,563 | 3,164 | 1,095 | 34.61% | 2,869 | 90.68% |
| INF | 2,857 | 2,728 | 515 | 18.87% | 2,593 | 95.05% |
| MUL | 2,754 | 2,351 | 1,945 | 82.73% | 1,853 | 78.82% |
| SS | 1,759 | 1,478 | 493 | 33.36% | 1,424 | 96.35% |
| MAT | 1,648 | 1,487 | 582 | 39.14% | 1,240 | 83.39% |
| AH | 184 | 98 | 17 | 17.35% | 89 | 90.82% |

# Figures

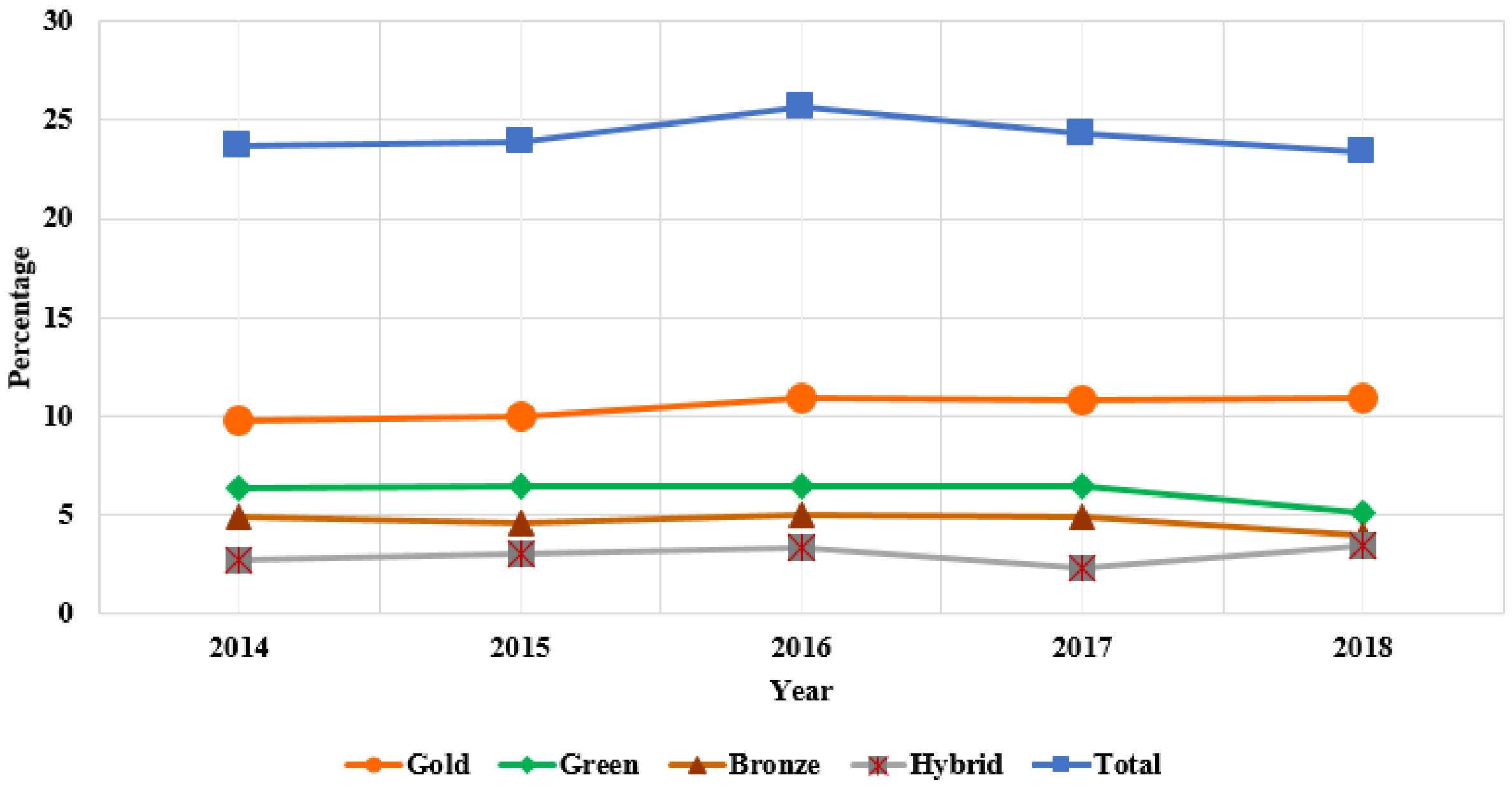

**Figure 1: Percentage of Articles that are Open Access, Plotted Year-wise**

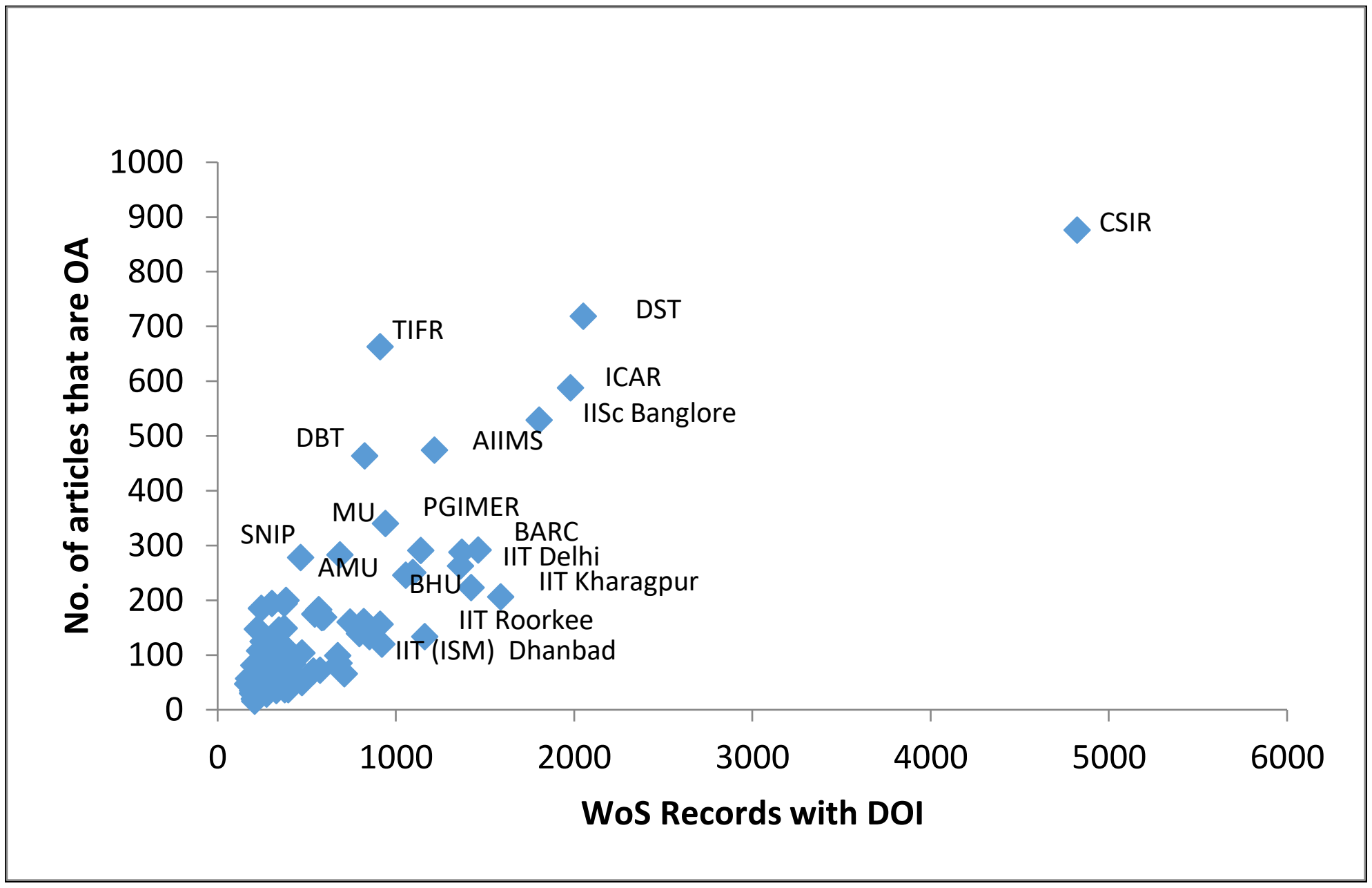

**Figure 2: Scatter plot of OA articles vs. total no. of articles in 100 most productive Indian Institutions**

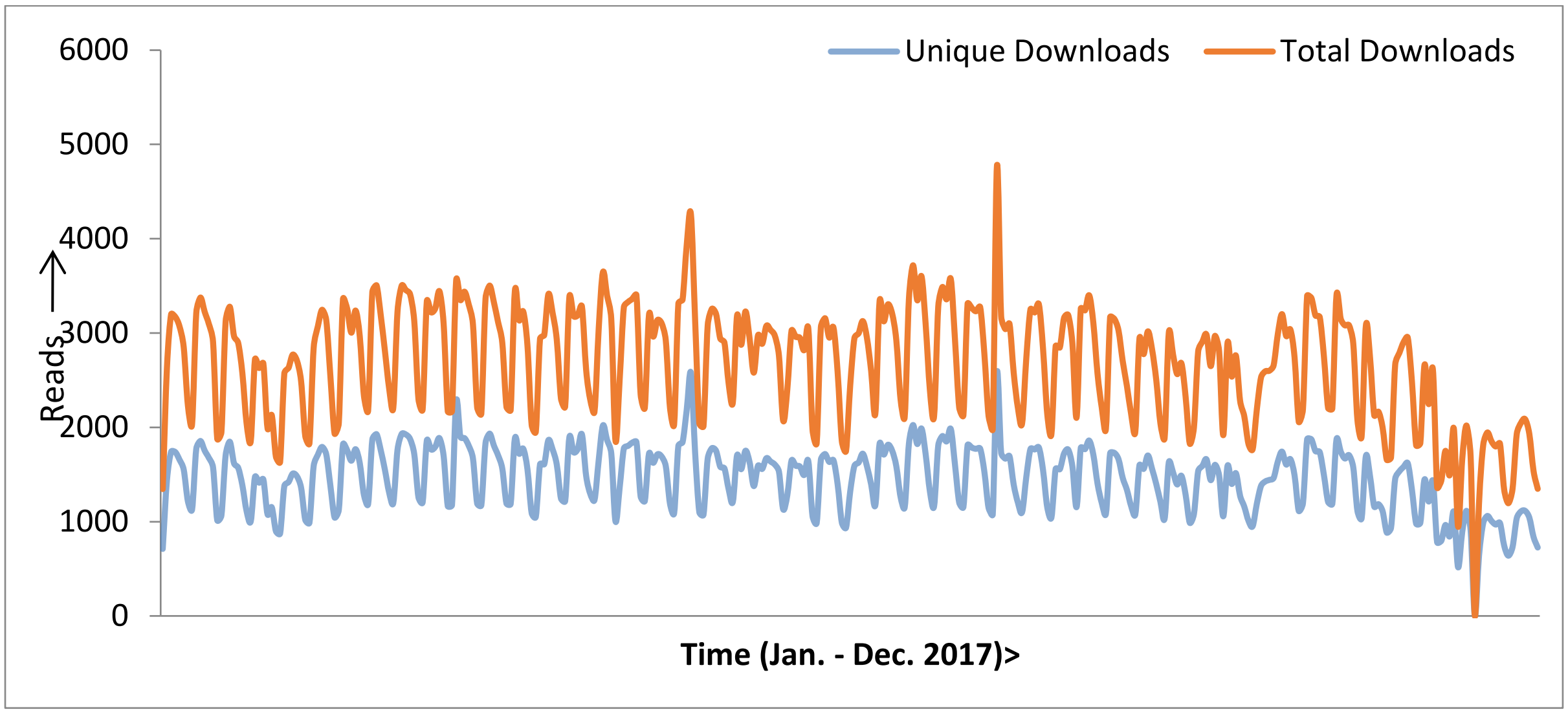

**Figure 3: Number of Indian Papers downloaded from Sci-hub (1.1.2017 to 31.12.2017)**

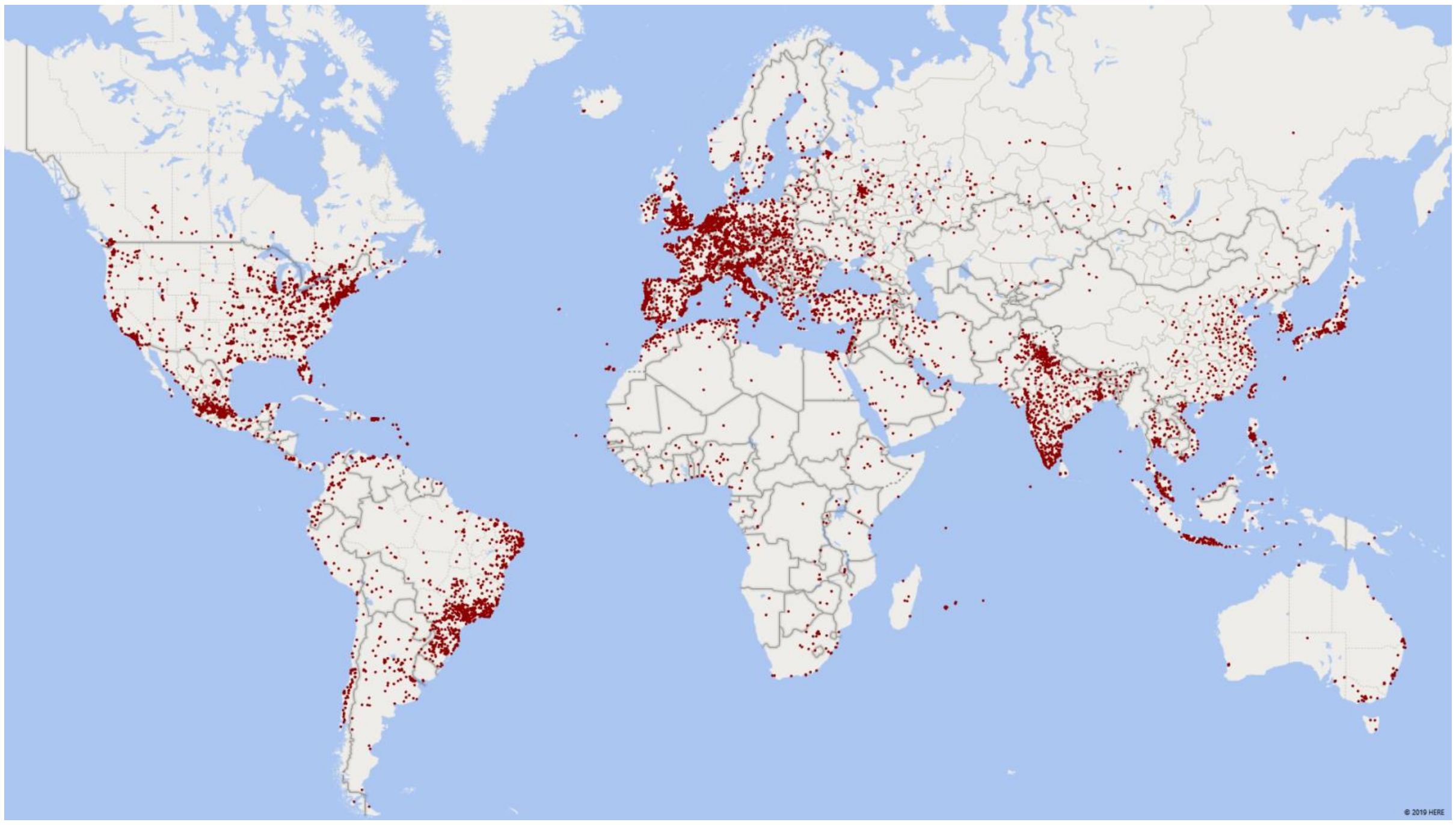

**Figure 4: Geographical distribution of download activity for Indian Papers in Sci-Hub**